\documentclass[twoside]{article}
\usepackage{fleqn,espcrc2}

\usepackage{graphicx}
\usepackage[figuresright]{rotating}

\newcommand{\AmS}{{\protect\the\textfont2
  A\kern-.1667em\lower.5ex\hbox{M}\kern-.125emS}}

\title{Origin of Neutrino Masses and Mixings}

\author{R. N. Mohapatra\address{Department of Physics, University of
Maryland, College Park, MD-20742,USA}%
        \thanks{This work is supported by a grant from the National
Science Foundation, grant no. PHY-9802552}}
       
\begin{document}

\begin{abstract}
A brief overview of the important theoretical issues in neutrino
physics is presented and some of the current ideas for addressing
them in the framework of the seesaw mechanism of neutrino
masses are discussed. Among the topics discussed are
 ways to understand maximal mixing
pattern among the three neutrino species, both uni-maximal and bimaximal
kinds. Constraints from baryogenesis on mixing patterns are
noted. In the last section, the question of understanding the
small masses for sterile neutrinos, which seem to be needed for various
reasons is addressed.
 \vspace{1pc}
\end{abstract}

\maketitle

\section{INTRODUCTION}
There is now strong evidence for neutrino masses and mixings
from the solar and atmospheric neutrino observations. The simplest way to
understand the deficits in neutrino fluxes observed in the above
experiments is to assume that the incident neutrinos oscillate into
another species which cannot be detected. For neutrino oscillations to
take place, they must have mass and mix
among themselves, with appropriate mass differences and mixing
angles. As far as the accelerator searches
for such oscillation effects go, with 
the exception of the Los Alamos experiment (LSND), all others
have yielded negative
results. These are of course not in contradiction with the solar and
atmospheric data since they probe different ranges of masses and mixings.
In fact, the negative results from the two experiments, CHOOZ and PALO
VERDE provide upper limits on one of the mixing angles that has
interesting interesting implications for theories of neutrino masses.
In this brief overview, I wish to draw attention to the theoretical
progress in understanding mass and mixing patterns in extensions of the
standard model, specifically the ones based on the seesaw mechanism that
seems to provide the simplest way to understand small neutrino
masses\cite{seesaw}. Due to space
limitations, this exposition will be brief, specially with respect to
references and we refer the reader to a recent review by the author on the
subject\cite{moh}.

\subsection{Four major theoretical issues in neutrino physics:}
The major issues of interest in neutrino theory are driven by the
following experimental results and conclusions derived from them. 
We will use the notation, where the flavor or weak
eigen states $\nu_{\alpha}$ (with $\alpha~=~e, {\mu}, {\tau}$)
are expressed in terms of the
mass eigenstates $\nu_{i}$ ($i=1, 2, 3)$ as $\nu_{\alpha} =~\sum_i
U^{MNS}_{\alpha i}\nu_i$. The $U^{MNS}_{\alpha i}$, the elements of the
Maki-Nakagawa-Sakata matrix represent the observable mixing
angles if the basis is such that the charged lepton masses are diagonal.
In any other basis, one has $U^{MNS}=U^{\dagger}_{\ell}U_{\nu}$, where the
matrices on the right hand side are the ones that diagonalize the charged
lepton and neutrino mass matrices. For simplicity, we will drop the
superscript MNS in what follows.

\subsubsection{Solar neutrinos:}
 There is no clear winner
among the various possible oscillation solutions to the solar neutrino 
puzzle\cite{concha}. In our discussion below, we will consider the small
angle MSW solution,
specially in the case when we want to understand the LSND results in the
framework of a four neutrino picture. Of particular interest now is
the large angle MSW solution, since it provides
one way to understand the reported flat energy distribution in the
data\cite{suzuki}. Vacuum oscillation requires very small mass differences
and we will not discuss it here, even though it also can a viable
possibility. Typical mass difference required is:
$\Delta m^2_{\nu_e-\nu_X}\sim 10^{-5}$ eV$^2$, with mixing angles either
maximal or small.

\subsubsection{ Atmospheric neutrinos:}
 Here evidence appears very
convincing that the likely explanation involves oscillation of $\nu_{\mu}$
to $\nu_{\tau}$, with $\Delta m^2_{\nu_{\mu}-\nu_{\tau}}\simeq 3\times
10^{-3}$ eV$^2$ and maximal mixing.

\subsubsection{ LSND:} The evidence for $\nu_{\mu}$ to $\nu_{e}$
oscillation from LSND needs to
be confirmed by another experiment. The KARMEN experiment has eliminated
part of the original LSND allowed domain. The MiniBOONE at Fermilab will
either confirm or refute these observations more definitively. But 
taking what we have at present seriously,
requires $\nu_{\mu}$ to $\nu_e$ mass difference
$\Delta m^2_{\nu_e-\nu_{\mu}}\simeq 0.2 -2$ eV$^2$ with points around
$\Delta m^2\simeq 6$ eV$^2$ also allowed with a small mixing (few percent 
level).

\subsubsection{ Neutrinoless double beta decay:} The most
stringent limits here are from the enriched $^{76}$Ge experiment by the
Heidelberg-Moscow collaboration and leads to a constraint on masses and
mixing angles as: $\sum_i U^2_{ei} m_i \leq 0.3 $ eV, with an uncertainty
of a factor of 2 due to nuclear matrix elements.

\subsubsection{ $U_{e3}$:} The reactor experiments CHOOZ and PALO
VERDE experiments imply that $U_{e3}\leq 0.16$.

The four major issues in neutrino physics can now be stated as follows:
\begin{itemize}

\item How does one understand the extreme smallness of the neutrino 
masses ?

\item How does one understand the maximal mixing angles among neutrinos
given that there is so much similarity between quarks and leptons and that
the quark mixings are small?

\item Could the neutrino masses (all or some) be degenerate ?

\item If LSND is confirmed, the only way to reconcile all data seems to be
to postulate the existence of a sterile neutrino\cite{calmoh}. The
question then is: how does one understand the ultralightness of
the sterile neutrino ? Note that since it is an electroweak singlet, its
mass could take any value.

\end{itemize}

\section{SMALL NEUTRINO MASSES}
It is well known that in the standard model the neutrino is massless due
to a combination of two reasons: (i) one, its righthanded partner
($\nu_R$) is absent and
(ii) the model has exact global $B-L$ symmetry. Clearly, to understand a
nonzero
neutrino mass, one must give up one of the above assumptions. If one
blindly included a $\nu_R$ to the standard model as a singlet, the status
of neutrino would be parallel to all other fermions in the model and one
would be hard put to understand why its mass is so much smaller than
that of other fermions. On the
other hand, instead of adding the $\nu_R$, one might exploit a prevailing
lore that gravity breaks all global
symmetries and include in the Lagrangian Planck mass suppressed higher
dimensional operators\cite{barbieri} of the form $LHLH/M_{P\ell}$
(where $L$ is a lepton doublet and $H$ is the Higgs doublet). These in
general lead to masses for
neutrinos of order $10^{-5}$ eV or less and are therefore not adequate
for understanding observations. Thus a nontrivial extension of the
standard model is called for.

\subsection{Does $m_{\nu}$ necessarily imply a right handed neutrino ?}
An interesting class of models were proposed in the 80's where one extends
the
stanadrd model without adding right handed neutrinos but extra Higgs
bosons to understand small $m_{\nu}$. In one class of models, one adds a
Higgs
triplet with $B-L=2$ with a large mass $M\gg M_W$\cite{ma}. The triplet
field acquires small vev of order $v_T\sim \frac{M^2_W}{M}$ via the type
II seesaw mechanism\cite{ma}. Thus, modulo the unknown origin of $M\gg
M_W$, this provides an understanding of the small $m_\nu$. There is
however one problem with this mechanism: to generate the baryon asymmetry
of the universe, one has to employ the triplet decay and out of
equilibrium
decay condition of Sakharov then implies that $M \geq 10^{13}$ GeV. This
scale is much higher than the conventional reheating temperature in
inflation models with supersymmetry\cite{sarkar}. Thus while this
mechanism may prove adequate for neutrino physics, it falls short of
expectations within a broader picture where one attempts a unifiied
understanding of different aspects of particle physics and cosmology.

A second approach is to attempt an understanding of small $m_\nu$ using
only electroweak scale physics and radiative corrections. There are
generic one loop \cite{zee} and 
two loop one\cite{babu} type models. The idea is to extend the standard
model by adding
$SU(2)_L$ singlet, electrically charged fields: $\eta (1,2)$ in the first
case and
$\eta^+$ and $h^{++}$ in the second case. The two loop mechanism has the
advantage that it requires less of a fine tuning to get the desired
smallness for $m_\nu$. However, both these approaches suffer from the
difficulty that
it is not easy to understand the origin of baryons in the universe. The
reason for that is that 
they necessarily involve $B-L$ violating interactions at the weak scale 
which will necessarily erase any baryons produced. 

\subsection{Right handed neutrino, seesaw mechanism and neutrino mass}
It seems rather compelling from the above reasoning that the small mass of
neutrinos provides a strong reason to believe in the existence of an
additional fermion state in nature i.e. the right handed neutrino, one per
generation. The inclusion of the $\nu_R$ expands the gauge symmetry of the 
electroweak interactions to $SU(2)_L\times SU(2)_R \times U(1)_{B-L}$
symmetry with fermion doublets $(u,d)_{L,R}$ and $(\nu, e)_{L,R}$ assigned
to the left-right gauge group in a parity symmetric manner; furthermore
the model has quark-lepton symmetry. More importantly, the elctric charge
formula for the model takes a very interesting form\cite{marshak}:
$Q= I_{3L} + I_{3R} + \frac{B-L}{2}$. It can be concluded from this that
below the scale $v_R$ where the $SU(2)_R\times U(1)_{B-L}$ symmetry
breaks down to the standard model and above the scale of $M_W$, one has
the relation $\Delta I_{3R}~=~-\Delta \frac{B-L}{2}$. This relation has
the profound consequence
that neutrino must be a Majorana particle and that there must be lepton
number violating interactions in nature.
In particular, the
$\nu_L-\nu_R$ mass matrix for three generations takes the form:
\begin{eqnarray} 
M~=~ \left(\begin{array}{cc} M_{LL} & M_{LR}\\ M^T_{LR} & M_{RR}
\end{array}\right)
\end{eqnarray}
where $M_{RR} = {\bf f}v_R$ is the Majorana mass matrix of the right
handed neutrinos, (${\bf f}$ is the new Yukawa coupling that determines
the right handed neutrino masses);
$M_{LL}~\simeq~{\bf f}\frac{ v^2_{wk}}{v_R}$ is the induced Majorana
mass matrix for the left handed neutrinos; $M_{LR}\equiv M_D = {\bf
Y}v_{wk}$ is the Dirac mass matrix connecting the left and the right
handed neutrinos. The diagonalization of this mass matrix leads to
following form for the light neutrino masses:
\begin{eqnarray}
M_{\nu}\simeq~{\bf f}\frac{ \lambda v^2_{wk}}{v_R}-\frac{1}{v_R} M^T_D{\bf
f}^{-1}M_D;
\end{eqnarray}
${\bf f}$, the Yukawa coupling matrix that is responsible for the mass
matrix of the heavy right handed neutrinos characterizes the
high scale physics, whereas all other parameters denote physics at the
weak scale. We have called this generalized formula for neutrino masses,
the type
II seesaw formula to distinguish it from the one that is commonly used in
literature where the
first term of Eq. (2) is absent. Important feature
of this formula is that both
terms vanish as $v_R\rightarrow \infty$ and since $v_R \gg v_{wk}$, the
smallness of the neutrino mass is much smaller than the charged fermion
masses. As was particularly emphasized in the third paper of
ref.\cite{seesaw}, the
dominance of V-A interaction in the low energy weak processes is now
connected to smallness of neutrino masses.

If in the above seesaw formula, the second term dominates, this leads to
the canonical type I seesaw formula and leads to the often discussed
hierarchical neutrino masses, which in the approximation of small mixings 
lead to $m_{\nu_{i}}\simeq m^2_{f_i}/v_R$, where $f_i$ is either a charged
lepton or a quark depending on the kind of model for neutrinos.

On the other hand, in models where the first term dominates, the neutrino
masses are almost generation independent. Therefore, if there is
indication for neutrinos being degenerate in mass \cite{cald} from
observations, one will have to resort to type II seesaw mechanism for its
understanding.

A further advantage of the right handed neutrino and seesaw mechanism is
that it fits in very nicely into grand unified frameworks based on
SO(10) models\cite{babu3}. The coupling constant unification then provides
a theoretical justification for the high seesaw scale and hemnce the small 
neutrino masses. Furthermore, the {\bf 16}-dim. spinor representation of
SO(10) has just the right quantum numbers to fit the $\nu_R$ in addition
to the standard model particles of each generation.

 \section{BARYOGENESIS CONSTRAINTS}
In the models with a heavy right handed neutrino, there is a definite
mechanism for baryogenesis via leptogenesis, which has been well discussed
in the literature\cite{fuku}. It is the decay of the right handed
neutrinos (denoted by $N_R$) $N_R\rightarrow \nu \ell$ that in conjunction
with the CP
violation in leptonic couplings generates lepton asymmetry\cite{buch}. The
lepton asymmetry is subsequently converted to baryon asymmetry via the
sphaleron interactions, associated with the electroweak physics. An
essential prerequisite for leptogenesis is that the interactions
involving the $N_R$ must be out of equilibrium i.e. $\Gamma_{N_R}\ll H$ (H
is
Hubble parameter at the relevant epoch). This equation roughly translates
into $\frac{h^2}{12\pi} \leq \frac{10 M_{N_R}}{M_{P\ell}}$, $h$ being the
Yukawa couplings of the relevant $N_R$, expected to be roughly of the same order
of magnitude as the quark or lepton Yukawa couplings. It is not very hard
to see that only the lightest of the three righthanded neutrinos (which in
a hierarchical picture is predominantly of electron type
i.e. $N_{eR}$) will have the
smallest coupling $h$ and will satisfy the out of equilibrium condition.
However since all the $N_R$'s are mixed among each other, the same
condition also implies that $N_{eR}$ mixing with $N_{\mu,\tau R}$
determined
by the ${\bf f}$ matrix better be very small. In quantitative terms, this
means that ${\bf f}_{e\mu}\leq \frac{m_e}{m_{\mu}}$ and $ {\bf
f}_{e\tau} \leq \frac{m_e}{m_{\tau}}$ or so. To see the kind of
constraints they imply on the neutrino masses and mixings, let us assume
type I seesaw and that in analogy with the quark sector, the mixings in
the Dirac neutrino mass matrix are negligible. One then gets:
\begin{eqnarray}
v^{-1}_R M^D_eM^D_{\mu}\sum_i U_{ei}U_{\mu i}m^{-1}_i \leq 5\times
10^{-3}\\
v^{-1}_R M^D_eM^D_{\tau}\sum_i U_{ei}U_{\tau i}m^{-1}_i \leq 2.5\times
10^{-4}.
\end{eqnarray}
These in turn can put severe constraints on models for neutrino masses. In
fact, there are
interesting examples neutrino mass models in the literature which have
difficulty satisfying Eq. (3).

\section{MAXIMAL MIXING IN GAUGE MODELS}
Let us consider only the three neutrino case, which is sufficient to
understand data if LSND results are not included. There are in general two
types mixing matrices that can fit data:
\noindent\underline{\it Case (i): Unimaximal mixing}
\begin{eqnarray}
U~=~\left(\begin{array}{ccc} 1 & \epsilon & 0\\
-\epsilon/\sqrt{2} & 1/\sqrt{2} & 1/\sqrt{2} \\
\epsilon/\sqrt{2} & -1/\sqrt{2} & 1/\sqrt{2} \end{array} \right).
\end{eqnarray}
\noindent{\underline{\it Case (ii): Bimaximal mixing\cite{bimax}}
\begin{eqnarray}
U~=~\left(\begin{array}{ccc} 1/\sqrt{2} & -1/\sqrt{2} & 0\\
1/{2} & 1/{2} & 1/\sqrt{2} \\
1/{2} & 1/{2} & -1/\sqrt{2} \end{array} \right)
\end{eqnarray}
Clearly, the neutrino mixing patterns described in the abive equations
bear no resemblace to that in the quark sector, in spite of the fact that
at the fundamental level there is complete quark lepton symmetry. The
difference must therefore be accounted for by whatever mechanism makes the
neutrinos Majorana particle e.g. seesaw mechanism. Below we explore
several ideas that are being explored in the literature to gain an insight
into the maximal mixing. Looking at the formula for
$U^{MNS}=U^{\dagger}_{\ell} U_{\nu}$, one can think of three ways:
(i) $U_{\ell}\simeq {\bf 1}$ but $U_{\nu}$ intrinsically different;
(ii) at the seesaw scale, both $U_{\ell,\nu}\simeq {\bf 1}$ but dynamics
of renormalization group extrapolation generates the maximal mixing;
(iii) $U_{\nu}\sim {\bf 1}$ but $U_{\ell}$ has the maximal mixing pattern. 
Let us briefly describe each one and discuss their relative merits.

\subsection{Maximal mixing due to $\nu_R$ texture}
From the type I seesaw formula, $M_{\nu}\simeq~- M^T_D{\bf
f}^{-1}M_D/v_R$, we see that this case is very generic to grand unified
theories where, one may expect to have $M_{\ell,\nu}\sim M_{q}$ so that
the entire mixing angle pattern arises from texture in the right handed
neutrino mass matrix. There is something natural about it since the origin
of the right handed neutrino mass matrix owes its origin to a different
source than the quark and lepton masses. There exist examples in
literature where this case is realized using symmetries or specific
ansatz for the right handed neutrino mass matrix\cite{sfk,akh}. 

\subsection{ Radiative magnification of mixing angles}
This is a very interesting where dynamics plays an essential role in
understanding the maximal mixings. The basic idea is that at the seesaw
scale, all mixings angles are small, a situation is quite natural if the
pattern of ${\bf f}$ Yukawa coupling is similar to the usual standard
model ones. Since the observed neutrino mixings are weak scale
observables, one must extrapolate\cite{babu1} the seesaw scale mass
matrices to the weak scale and recalculate the mixing angles. 

The extrapolation formula is 
\begin{eqnarray}
{\cal M}_{\nu}(M_Z)~=~ {\bf I}{\cal M_{\nu}} (v_R) {\bf I}\\
where~~~~~~~~{\bf I}_{\alpha \alpha}~=~
\left(1-\frac{h^2_{\alpha}}{16\pi^2}\right)
\end{eqnarray}
Note that since $h_{\alpha}= \sqrt{2}m_{\alpha}/v_{wk}$ ($\alpha$ being
the charged lepton index), in the extrapolation only the $\tau$-lepton
makes a difference. In the MSSM, this increases the ${\cal M}_{\tau\tau}$
entry of the neutrino mass matrix and essentially leaves the others
unchanged. It was shown
in a recent paper\cite{balaji} that if the muon and the tau neutrinos are
nearly degenerate in mass at the seesaw scale, and in supersymmetric
theories, the $tan\beta\geq
5$, the radiative corrections can become large enough so that at the weak
scale the two diagonal elements of ${\cal M}_{\nu}$ which were nearly
equal but different at the seesaw
scale become so degenerate that the mixing angle becomes maximal and a
solution to the atmospheric neutrino deficit emerges
even though at the seesaw scale, the mixing angles were small. 
This happens only if the experimental observable $\Delta m^2_{23}\leq 0$
i.e. the usual hierarchical pattern. This possibility can be tested in
experiments such as the ones being contemplated in future neutrino
factories. This mechanism will become more compelling if the LSND
results are confirmed since in that case the $\nu_{\mu}$ and $\nu_{\tau}$
are nearly degenerate or evidence for a small neutrino hot dark matter
component emerges.

An interesting question is whether this mechanism can be extaended
to the case of three generations and whether it can explain the
bimaximal pattern also. This question was investigated in the
ref.\cite{balaji} and it was found that the answer to the first
question is yes and to the second, it is  ``no''. Furthermore for the
first case to work, one must explain the solar neutrino
puzzle by the small angle MSW mechanism if one has to be consistent with
the CHOOZ-PALO-VERDE constraint.

\subsection{ Lopsided unification}
This basic idea of this class of models is that in the seesaw formula,
$M^D$ and $M_{RR}$ are generic with very little generation mixing but
$M_{\ell}$ is
special. The reason this is consistent with grand unification of quarks
and leptons is that in some GUT theories such as $SU(5)$, one has the
relation $M_{d}~=~M^T_{\ell}$. What this means is that the lepton
mixings are related to mixings among the right handed
down quarks, which are not observable and thus need not be small. As a
result, the
$U_{\ell}$ could harbor large mixing angles without
contradicting quark-lepton unification. If fermion masses
are asymmetric as in SU(5) theories or in $SO(10)$ theories with {\bf 120}
type Higgs bosons, then one could realize this
scheme\cite{albright}. Of course, the largeness of the mixing angles must
be put in by hand.

\subsection{Bimaximal mixing}
In this subsection, let us briefly discuss the bimaximal mixing and their
theoretical origin. It has been realized\cite{nuss} that the neutrino mass
matrix that leads to bimaximal mixing pattern is:
\begin{eqnarray}
{\cal M}_{\nu}~=~ \left(\begin{array}{ccc} \delta & F & F \\
F & A & -A \\
F & -A & A \end{array} \right)
\end{eqnarray}
There are two subcases of this mass matrix with very different
experimental tests; case (i) $\delta \ll A \ll F$: In this case, the
neutrino mass hierarchy is inverted i.e. $\Delta m^2_{23}\geq 0$. On the
other hand, there is another case, i.e. case (ii) where $\delta \ll F \ll
A$. In this case, the mass pattern has normal hierarchy i.e. $\Delta
m^2_{23} \leq 0$. Clearly, experiments such as the ones contemplated using
long baseline using nutrino factories can decide between these two very
important cases.

As far as the theoretical origin of these mass patterns is concerned,
several interesting directions have been pursued. One way which quite
appealing is to use a $L_e-L_{\mu}-L_{\tau}$ symmetry to restrict the
mass matrix\cite{anjan}. This is however a weak scale theory and does not
possess the advantages of seesaw type models. There have been attempts to
derive these patterns from grand unified theories\cite{chen}.

\section{Sterile neutrinos}

If the existence of the sterile neutrino\cite{bil} becomes
confirmed say, by a confirmation of the LSND observation of
$\nu_{\mu}-\nu_e$
oscillation or directly by SNO neutral current data, a key theoretical
challenge will be to
construct an underlying theory that embeds the
sterile neutrino along with the others with appropriate mixing pattern,
while naturally explaining its ultralightness.

If a sterile neutrino was introduced into the standard
model, the gauge symmetry would not forbid its bare mass, implying that
there would be no reason for it mass to be small. It is a common
experience in
physics that if a particle has mass lighter than normally expected on the
basis of known symmetries, then it is an indication for the existence of
newsymmetries. This line of reasoning has been pursued in recent
literature
to understand the ultralightness of the sterile neutrino by using new
symmetries beyond the standard model.

There are several suggestions for this new symmetry that might help us to
accomodate an ultralight $\nu_s$ and in the process lead us to new
classes of extensions of physics beyond the standard model. Two examples
are: (i) one based on the $E_6$ grand unification model, where the {\bf
27}-dim. representation which is used to represent the quarks and leptons
has a standard model singlet neutrino, which could then be a candidate for
the sterile neutrino and (ii) another based on the possible existence of
mirror matter in the universe, that we discuss later.

A completely different way to understand the ultralightness of the sterile
neutrino is to introduce large extra dimensions in a Kaluza-Klein
framework and not rely on any symmetries. In these models, the
sterile neutrino is supposed to live in the bulk; therefore if the bulk
size is of order milli-eV$^{-1}$ as is generally contemplated, the sterile
neutrino can be identified as one of the KK excitations of a massless bulk
neutrino. For a discussion of this approach, see \cite{dienes}.

Let us now briefly touch on the mirror universe model for the
ultralight neutrino\cite{bere,foot} since the idea has implications
far beyond the domain of neutrino physics. The basic idea
is that there is a complete duplication of matter and forces in the
universe (i.e. two sectors in the universe with matter and gauge forces
identical prior to symmetry breaking) as suggested , say by the string
theories.
The mirror sector of the model will then have three light neutrinos
which will not couple to the Z-boson and would not therefore have been
seen at LEP. Denoting the fields in the mirror sector by a prime over
the standard model fields, the $\nu'_i$ can qualify as the sterile
neutrinos of which
we now have three. The lightness of $\nu'_i$ is dictated by the mirror
$B'-L'$ symmetry in a manner parallel to what happens in the standard
model. Thus the ultralightness of the sterile neutrinos is understood in
a most ``painless'' way.

The two ``universes'' are assumed to communicate only via
gravity or other forces that are
equally weak. This leads to a mixing between the neutrinos of the two
universes and can cause oscillations between $\nu_e$ of our
universe to $\nu'_e$ of the parallel one in order to explain for example
the solar neutrino deficit. The atmospheric neutrino deficit in this case
can be explained by using the usual $\nu_{\mu}-\nu_{\tau}$ oscillation.

There are two versions of this model: one where the masses in both sectors
are identical for corresponding particles\cite{foot}. This is called the
symmetric miror model. On the other hand, one can envision a scenario
where after symmetry breaking all masses scale by a common
factor\cite{teplitz}. The second one has certain cosmological advantages,
that mirror matter can become the dark matter of the universe.

\section{CONCLUSION}
In conclusion, the seesaw mechanism appears by far to be the
simplest way to understand the small neutrino masses.
A grand unified framework such as the one based on the SO(10) can then
help us to understand the high seesaw scale as a consequence of
gauge coupling unification. This also
provides a way to understand the origin of baryons in the universe. Our
understanding of mixings on the other hand, is
at a preliminary level. A
particular challenge to theorists is to understand the so called bimaximal
mixing pattern, which will emerge as the favorite, if the solar neutrino
deficit is to be solved via the large angle MSW or the vacuum oscillation.
Finally we comment on the status of the sterile neutrino and understanding 
of its tiny mass in a ``natural'' framework.

\end{document}